\documentclass[12pt]{article}
\usepackage{graphicx}
\newcommand{\figwidth}{6in}
\begin{document}

\title{Combining the ApEn statistic with surrogate data analysis for the
detection of nonlinear dynamics in time series}

\author{Randall A. LaViolette,\thanks{Mailstop: 2208, e-mail: yaq@inel.gov}
\ Charles R. Tolle,\\ Timothy R. McJunkin and Daphne L. Stoner\\
\textsc{idaho national engineering \& environmental
laboratory}\thanks{Operated for the U. S. Department of Energy
under DOE-ID Operations Office Contract DE-AC07-99ID13727.} \\
\textsc{p.o. box 1625, idaho falls id, 83415, usa} }

\maketitle

\begin{abstract}
We tested the natural combination of surrogate data analysis with
the ApEn regularity statistic developed by Pincus [\emph{Proc.
Natl. Acad. Sci. USA} \textbf{88} (1991) 2297] by applying it to
some popular models of nonlinear dynamics and publicly available
experimental time series. We found that this easily implemented
combination provided a useful method for discriminating signals
governed by nonlinear dynamics from those governed by linear
dynamics and noise. An apparently novel physical interpretation
of ApEn also is supplied.
\end{abstract}

\section{Introduction}
Significant developments in the last decade have provided
techniques that can discriminate between signals generated by
essentially linear dynamics and signals generated by essentially
nonlinear and possibly chaotic
dynamics\cite{gollub}\cite{thermo}\cite{glass}\cite{abarbanel}\cite{kantz}.
Our goal here is to assess the utility of the combination of
surrogate data techniques with the ``ApEn'' (or ``approximate
entropy'') statistic for the identification of essentially
nonlinear signals. In the next two subsections we briefly
recapitulate surrogate data methodology and the ApEn statistic,
respectively. We present the results of applying the combined
analysis to various data sets and models of nonlinear dynamics in
the next Section. We conclude in Section 3 with some comments on
the results. Three short appendices follow that set forth some
details of the algorithm for calculating ApEn, the connection
between ApEn and standard quantities, and the previously
unnoticed numerical coincidence of ApEn with the thermodynamic
entropy of the one-dimensional Ising model, respectively.

\subsection{Surrogate Data Methodology}
The idea of surrogate data is that one tests some hypothesis
concerning a time series by comparing the value of some statistic
$Q$ for the original time series with the sample mean of that
statistic $\bar{Q}$ in an ensemble of random replicates (the
surrogates) generated from the original time series under that
hypothesis. The hypothesis would be taken to be true if the
original time series were statistically indistinguishable from
the ensemble of surrogate time series. ``The use of surrogate
data is essential for deciding whether an irregular time series
arises from nonlinear deterministic chaos or linear stochastic
dynamics. The method was introduced by Theiler et al.
(1992).''\cite{glass} The particular method that we are concerned
with was designated FT by Theiler et al.
\cite{theiler}\cite{kennel}\cite{prichard}, and tests the
following linearity hypothesis: The original time series is the
result of linear stationary dynamics driven by gaussian white
noise, i.e., the data can be modeled by ARMA
dynamics\cite{kantz}. The surrogates are produced by first
randomizing the phase of the Fourier transform of the original
time series, then inverting back to the time domain to create a
new time series in the ensemble of surrogate time series. This
procedure ensures that each surrogate time series possesses the
same power spectrum as the original data. Theiler et al.
suggested that the ``significance'', i.e., the number $S$ of
standard deviations $\sigma_{\bar{Q}}$ of $\bar{Q}$ between $Q$
and $\bar{Q}$, written as
\begin{equation}
S = \frac{|Q - \bar{Q}|}{\sigma_{\bar{Q}}}, \label{significance}
\end{equation}
could be employed as a simple albeit crude test for the success or
violation of the linearity hypothesis. They also suggested that
the statistic $Q$ could be the correlation integral or
correlation dimension, suggestions that have been widely adopted,
although other statistics also have been suggested for the
surrogate data
method\cite{eeg}\cite{reverse}\cite{palus}\cite{deco}\cite{kocarev}.
In view of the usual uncertainties in the estimate of
$\sigma_{\bar{Q}}$ and the crudeness of $S$ itself\cite{theiler},
in this work we required $S \geq 10$ to score a violation,
although other authors\cite{kocarev} have been willing to declare
a violation of the linearity hypothesis with a significance as
small as three. The analysis is meaningful only for stationary
time series; examples of violations of the linearity hypothesis
by linear non-stationary processes have been
documented\cite{timmer}.

Several authors have proposed modifications to the FT method for
generating surrogate series. Theiler et al. themselves introduced
the ``amplitude-adjusted'' method (designated AAFT by
them\cite{theiler}) because the amplitudes of the surrogate data
generated by the FT method can vary significantly from the
envelope of the original time series. It turned out that the AAFT
itself introduced other concerns that in turn have been addressed
by yet more elaborate and in some cases more computationally
demanding surrogate data
methods\cite{kantz}\cite{schreiber}\cite{kugiumtzis}. Those
proposals to modify the FT method retained the original
suggestions for the statistics; here we pursue the opposite
approach, retaining FT here for computational and conceptual
simplicity, and employing ApEn as an alternative statistic. Of
course ApEn could have been just as easily employed with any of
the other methods to generate a surrogate series.

\subsection{ApEn: A regularity statistic}
In this paper we address only the choice of the statistic in the
FT method. One of our concerns with both the correlation integral
and the correlation dimension as a statistic is that each may
require a large number of data points. The data required may grow
rapidly with the correlation dimension, so that some workers have
had to employ this or related statistics in a regime where the
validity of doing so was difficult to
assess\cite{theiler86}\cite{eckmann}. One of the appeals of ApEn
is that it has performed well, in other
contexts\cite{pincus91}\cite{pincus92}\cite{pincus95a}\cite{pincus95b}\cite{pincus97}\cite{chatterjee},
even on decimal data sets as small as $N \approx 100$, or binary
data sets as small as $N \approx 20$. Therefore we investigated
the application of ApEn to the FT surrogate data method, which we
designate ApEn+FT.

The statistic ApEn was developed by Pincus to measure the
conditional probability that ``...runs of patterns that are close
for [some number] of observations remain
close...''\cite{pincus91}. ApEn therefore depends upon the length
($N$) of the series, the width ($m$) of the window that defines
the patterns , and the tolerance ($r$) that defines closeness of
the patterns, and incidentally provides ApEn with some resistance
to noise. Pincus' algorithm for the calculation of ApEn is
recapitulated in the Appendix A for reference. The theoretical
(large $N$, $m$ and small $r$) bounds to ApEn are zero for a
perfectly regular sequence, and $\ln(B)$ for a maximally
irregular (i.e., random) sequence of base $B$ numbers. It turns
out (see Appendix C) that ApEn numerically coincides with the
thermodynamic entropy for the one-dimensional Ising model when
each configuration of spins is interpreted as a string of bits.
This coincidence serves at least to provide a physical
interpretation of ApEn (for binary data), even though ApEn was
constructed without any explicit reference to statistical
mechanics or thermodynamics. However, we did not further exploit
this connection. Furthermore, in spite of its name and its
asymptotic behavior (see Appendix B), we did not regard ApEn as an
approximation to the, e.g., Kolmogorov-Sinai or Eckmann-Ruelle,
non-thermodynamic entropies because the amount of data required
to achieve useful approximations may be very large in some
circumstances\cite{theiler86}\cite{eckmann}\cite{pincus91}, and
because those entropies can be especially sensitive to noise. We
also did not compare ApEn with either the correlation integral or
correlation dimension for three reasons. First, we have not
attempted to establish the superiority of ApEn over all other
statistics in all circumstances. Second, both the correlation
integral and the correlation dimension nominally require the
evaluation of limits that in turn may also require much larger
quantities of data than ApEn itself\cite{pincus91} (see also
Appendix B). Third, the comparison between ApEn and the
correlation integral (for example) is complicated by the fact
that the choice of parameters $m$ and $r$ that are optimal for
one are almost certainly not optimal for the other; employing
only the parameters that we also employed for the ApEn
calculation might have appeared to be creating a ``straw man''.

Therefore, we adopted the alternative recommended by
Pincus\cite{pincus95a}, i.e., that ApEn is simply a regularity
statistic (or family of statistics) defined by the choice of its
($N,m,r$) parameters that stands on its own. As such, ApEn has
been applied to binary sequences\cite{pincus97}\cite{chatterjee}
and to physiological
data\cite{pincus94}\cite{cardiac}\cite{heart}. Although ApEn is a
biased statistic, as are many nonlinear statistics, we always
employed it to compare only the sequences with the same set of
($N,m,r$) parameters, so that the bias would not become an issue
here. ApEn is strongly sensitive to the complexity of the data,
but is insensitive to other attributes, as would be desired of
such a statistic. For example, and not surprisingly, ApEn is
insensitive to topological conjugation\cite{thermo} of sequences.
Furthermore, ApEn is semi-pivotal in the sense that it is always
insensitive to the sample mean of the data, and it is insensitive
to the sample variance if, as Pincus suggests\cite{pincus91}, the
tolerance parameter $r$ is always chosen as a fixed fraction of
the sample standard deviation. This feature contributed to the
utility of ApEn in surrogate data methodology\cite{prichard}; in
particular, the concern that inspired the AAFT surrogate data
method should be insignificant here.

Even though the combination of ApEn with surrogate data
methodology is a natural one to employ, we found no examples of
such analysis in the physical science or engineering literature.
However we recently discovered two examples of this kind of
analysis in the cardiology literature\cite{cardiac}\cite{heart},
although those applications appeared to us to be narrow, with no
attempt to test the method with standard data or models, or to
explore its limitations. Therefore we proceed to show the results
of our tests of ApEn+FT in the next section.

\section{Results of tests with ApEn+FT}
The plan of this section consists of the employment and empirical
testing of the ApEn statistic in the FT surrogate data method
(ApEn+FT) to both experimental and computational data. In order
to obtain consistent results, we followed published
recommendations\cite{pincus91}\cite{pincus95a} to fix the pattern
window length to $m=2$ and the tolerance to $r = 0.2\sigma$,
where $\sigma$ was the sample standard deviation estimated by the
non-parametric bootstrap\cite{efron}. Although important
questions remain concerning the optimal values of $N$, $m$, and
$r$, we make no pretense of answering them here but defer them to
future work. However, preliminary estimates (for decimal numbers)
suggest that the minimum number of points for an adequate
calculation of ApEn requires on average only about $30^m$
points\cite{pincus95a}. Therefore we always considered data sets
at least as large as $N = 2048$. On the other hand, we did not
examine data sets larger than $N = 8192$ ($N = 4096$ for most
data sets), because we wanted to see if the ApEn+FT analysis
would be useful for medium-size data sets; for very large data
sets it is conceivable that any number of methods would work
equally well, but such large data sets are rarely achieved
experimentally. We also restricted $N$ to powers of two so that
we could use the simplest version of the Fast Fourier
Transform\cite{recipes} without having to pad the data.

For each original time series we generated an ensemble of only
ten FT surrogate time series, because the fluctuation of ApEn in
the ensemble was consistently small and because occasional tests
with 40 surrogates did not yield significantly different
estimates. As discussed below, the length $N$ of the series had a
much greater impact on the performance of the hypothesis test
than did the size of the surrogate ensemble. The percentile
bootstrap method was employed to estimate the (0.5\%,99.5\%) or
(0.1\%,99.9\%) confidence intervals employed in the Figures in
order to assist visual inspection, but only the bootstrap
estimate of the standard deviation was employed in the
significance calculations (see Equation
\ref{significance})\cite{efron}. The bootstrap usually provides a
good estimate of the sample standard deviation even for ten
samples, especially where, as was the case here, the median and
the mean are nearly coincident.

We did not display the time series themselves because we found
that it was misleading to visually compare the original time
series with its surrogates. The surrogate data do not need to
overlap point for point with the original in order for them to be
statistically the same. The FT method randomizes the phase, but
ApEn is properly insensitive to the phase (if any) present in the
data, so that even when the original and surrogates have the same
ApEn, the time series themselves may appear to be quite different
from one another. On the other hand, some series (e.g.,
H\'{e}non) vary so rapidly, that the original and surrogates may
appear similar when they are actually statistically distinct.

\subsection{Noise}
We began by studying the behavior of ApEn for white noise, in
order to provide a context for the values of ApEn for the
subsequent data, and to motivate the combination of ApEn with the
surrogate data techniques. We generated an ensemble of ten time
series from a white noise source provided here by two uniform
pseudorandom number generators, respectively: a standard
multiplicative-congruential generator, implemented as
``ran2''\cite{recipes}, and a much different generator employing a
random-walk algorithm implemented as ``rawkrab''\cite{rawkrab}.
Each generator gave the same result for the sample mean, i.e.,
$\mathrm{ApEn}(N=4096$, $m=2$, $r = 0.2\sigma) = 2.11$ (with a
standard deviation estimated to be 0.01), about $9\%$ below the
theoretical maximum $\mathrm{ApEn}( N\rightarrow\infty,
m\rightarrow\infty, r\rightarrow 0)=\ln(10)$\cite{pincus91}. We
continued this calibration by generating another ensemble of ten
time series directly from the Brownian noise (or
Ornstein-Uhlenbeck) power spectrum\cite{rice}
\begin{equation}
S_0(\omega,\gamma) = \frac{2\gamma\sqrt{N}}{\pi(\omega^2 +
\gamma^2)}, \label{noise}
\end{equation}
for each dissipation rate $\gamma$, i.e., we assigned random
phases to $S_0$ and computed its inverse Fourier transform to
obtain each trajectory\cite{osborne}. Such trajectories
automatically satisfy the linearity hypothesis. The resulting
time series is a solution of a Langevin equation, which can be
represented simply for discrete times by the ARMA
model\cite{gaspard} $x_n = c_0 + c_1 x_{n-1} + c_2 \xi_n$, with
time-invariant coefficients $c_j(N,\gamma)$, and the uncorrelated
gaussian noise of zero mean and unit variance $\xi_n$.
\begin{figure}
\centerline{\includegraphics[width =
\figwidth]{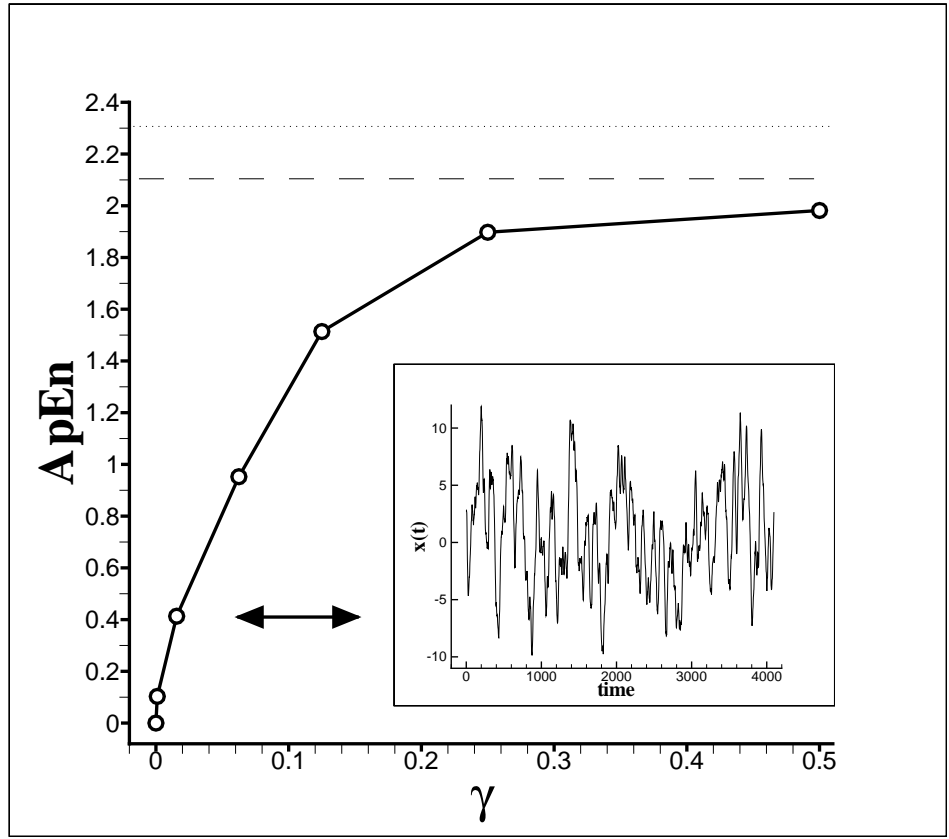}}\caption{ApEn for white and colored noise.
Each circle corresponds to sample mean of ApEn computed for an
ensemble of ten trajectories generated (via uniformly random
phases) from the Brownian noise spectrum $S_0(\omega,\gamma)$
(Equation(\ref{noise})), for various values of $\gamma$. The
standard deviation is estimated to be 0.01, about $1/10$ the
diameter of the circles. The inset shows a typical time series
reconstructed from $S_0(\omega,\gamma = 1/64)$. For reference, the
dashed line shows the value of ApEn uniformly distributed
(pseudo)random numbers on the interval $[0,1]$. The dotted line
shows the value of $\ln(10)$, the ApEn for uniform random decimal
numbers in the limit of large $N$ and $m$ and small $r$. The
circle at the origin is theoretical, not computed.}
\label{fig-noise}
\end{figure}
Figure \ref{fig-noise} shows the sample mean ApEn computed for
each of the various $\gamma$; the standard deviation (estimated
by bootstrap techniques\cite{efron}) is about $0.01$, or about
$1/10$ the diameter of the circles. For small $\gamma$, the
spectrum becomes a smeared delta function, the trajectories
become more regular, and ApEn is near zero. For larger $\gamma$,
the spectrum becomes flatter, the trajectories become more
random, and ApEn approaches its white noise limit (i.e., $2.11$)
for the parameters employed above. For intermediate values of
$\gamma$ the trajectories appear noisy but more regular, as shown
in the inset of Figure \ref{fig-noise} for $\gamma = 1/64$. Thus
any allowed value for ApEn can be found from the Brownian time
series. Therefore, instead of relying upon the value of ApEn
alone, we combined it with surrogate data generation in order to
test the hypothesis of linear dynamics.

\subsection{Data sets and models}
\begin{figure}
\centerline{\includegraphics[width =
\figwidth]{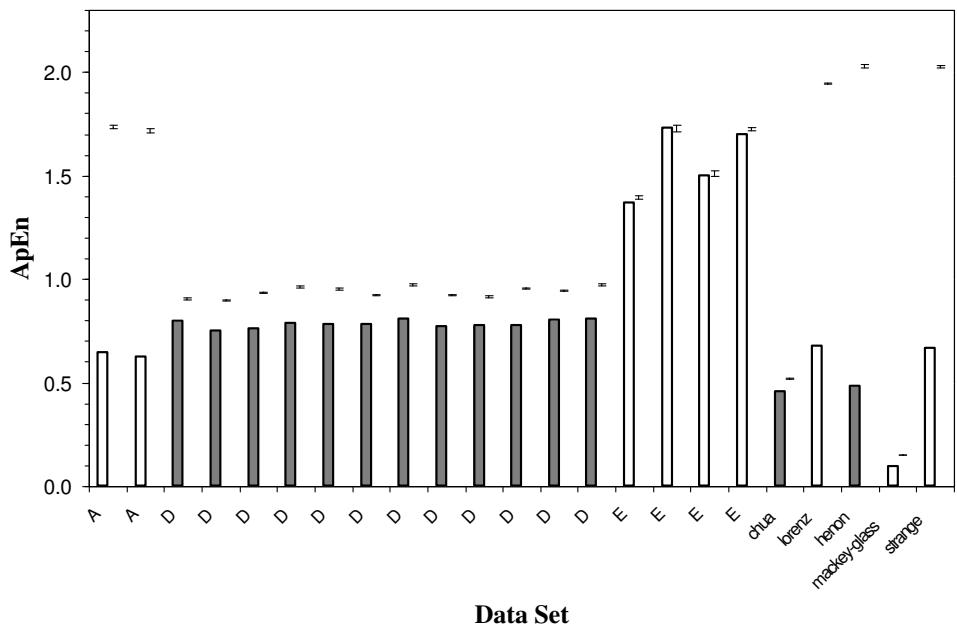}}
\caption{ApEn of original and surrogate time
series for various data sets and models. The shading of the
columns is used here to group data sets. The top of each column
shows the ApEn of the original time series. The height of each
pair of error bars shows the $(0.5\%,99.5\%)$ bootstrap confidence
interval about sample mean (not shown) of the ApEn of the
surrogate time series.  The labels ``A'', ``D'', and ``E''
correspond to the Santa Fe Institute 1991-1992 Competition Data
Sets A:continuation, D and E, respectively. Multiple columns
labeled A and D correspond to subsets split in increments of 4096
points from the original data set. The columns labeled E
correspond to parts 8, 11, 13, and 14, respectively, of Data Set
E, and contain only the first 2048 measurements of each part.  The
column labeled ``chua'' corresponds to the first 4096 points of a
relatively noiseless 5000-point time series of the voltage from an
experimentally realized network of two Chua circuits. The next two
columns correspond to $N = 8192$ time series generated numerically
by us for the x-coordinate of the Lorenz system ($Pr = 16$, $Ra =
45.92$, and $b = 6$) and the x-coordinate of the $N = 4096$
H\'{e}non map ($a=1.4$ and $b=0.3$), respectively. The next column
corresponds to the $N=8192$ time series generated numerically by
us for the Mackey-Glass model. The last column corresponds to the
$N = 4096$ time series generated numerically by us for the
x-coordinate of the two-dimensional non-chaotic strange attractor
($\lambda = 1.5$) of Grebogi et al.} \label{fig-data}
\end{figure}

We applied ApEn+FT to data sets and models that have been widely
discussed and for which the time series can be either readily
computed or freely downloaded, as follows: the data sets labeled
``A'' (chaotic laser), ``D'' (turbulent flow), and ``E'' (light
variability from a star) were taken from the Santa Fe Institute's
1991-1992 competition\cite{weigend}\cite{santafe} (but not the
other data sets that were represented as significantly
non-stationary); an experimental realization of the Chua
circuit\cite{chua}\cite{chua_dat}; the x-coordinate of the Lorenz
system\cite{abarbanel}; the x-coordinate of the H\'{e}non
map\cite{thermo}; the Mackey-Glass model\cite{glass}; and the
x-coordinate of the two-dimensional map of the non-chaotic strange
attractor of Grebogi, Ott, Pelikan, and Yorke\cite{ott}. In all
we obtained 23 original time series, for which we first computed
ApEn, as shown in Figure \ref{fig-data}. For each of these series
we also generated an ensemble of ten FT surrogate time series, and
calculated the sample-mean ApEn for each ensemble. We calculated
$(0.5\%,99.5\%)$ confidence intervals about by the non-parametric
percentile bootstrap method\cite{efron}\cite{boot}. The results
are displayed in Figure \ref{fig-data} and Table
\ref{tab-significance}.
\begin{table}[h]
\caption{Significance of ApEn+FT test} \vspace{12pt}
\centerline{\begin{tabular}{|c|c|c|} \hline
Data Set & $N$ & $S$\\
\hline
A (mean)& 4096 & 120\\
D (mean)& 4096 & 50 \\
E (all) & 2048 & $< 5$  \\
chua & 4096 & 30\\
lorenz & 8192 & 300 \\
henon & 4096 & 200 \\
mackey-glass & 4096 & 5 \\
mackey-glass & 8192 & 25 \\
strange & 4096 & 230 \\
\hline
\end{tabular}}
\label{tab-significance}
\end{table}

Both visual inspection of Figure \ref{fig-data}, and the
estimation (see Equation \ref{significance}) of the significance
$S$ (Table \ref{tab-significance}) indicated that the series
labeled \emph{A, D, chua, lorenz, henon, mackey-glass,} and
\emph{strange} clearly violate the linearity hypothesis, as
expected. To put these results into perspective,
others\cite{kocarev} have been willing to declare a violation of
the linearity hypothesis with a significance as small as $3$
(using other statistics), while we insisted upon a significance
of at least $10$ (see Equation \ref{significance}). The small
magnitude of the fluctuations of ApEn for the surrogate data
contributes substantially to the large values of $S$ in the case
of violations of the linearity hypothesis. The Mackey-Glass system
is the smoothest and most nearly periodic of any of the time
series that we considered here and was also one of the most
resistant to our analysis. The time series of 4096 points (not
shown in Figure \ref{fig-data}) resulted in a significance well
below our criterion of $S \geq 10$. In order to find a more
significant violation of the linearity hypothesis, we generated
the 8192-point time series (see Table \ref{tab-significance}).
Again, the fluctuations in the ApEn of the surrogate data were
small. On the other hand, the Lorenz system did not need $N =
8192$ for the violation to become apparent with this analysis; we
presented the results in order to indicate the extreme
significance that could be achieved with this analysis.

The ApEn+FT analysis indicated that series E (light variability
of a star) did satisfy the linearity hypothesis, as suggested by
both visual inspection of Figure \ref{fig-data} and estimation of
the significance (Table \ref{tab-significance}). If E were
actually governed by nonlinear dynamics, then this would have
been a failure of our analysis. Our difficulty with the
Mackey-Glass system also suggests that our analysis might have
failed, because at $N = 2048$, the series in E are the shortest
that we have considered.  On the other hand, inspection of the
data sets themselves and their power spectra suggested that the
fluctuations in E might be more random than recent and more
substantiated examples of nonlinear light variability of
astronomical objects\cite{leighly}\cite{timmer00}. Therefore we
were unable to decide between the conclusion that the data in E
really are nonlinear but are obscured (to our analysis) by noise,
or that the data in E really are governed by ARMA dynamics, as
indicated by our analysis.

\subsection{Sensitivity to noise}
In order to gain a sense of the magnitude of the noise needed to
disrupt our analysis, we added uncorrelated gaussian noise
$\xi(t)$ (zero mean, unit variance), scaled by a factor, to the
experimental Chua circuit and the H\'{e}non map. We chose these
time series for comparison because the Chua system violated the
linearity hypothesis with relatively small significance ($S
\approx 30$) while the H\'{e}non map had a much larger
significance ($S \approx 200$). By trial and error we estimated
the scale factor $\phi$ such that the addition of $\phi\xi(t)$ to
the time series reduced $S$ to ten. In Table \ref{tab-reduce} we
show the range (i.e., distance between the extreme values) of the
original time series, $\phi$, and $\rho$, the corresponding
signal-to-noise ratio\cite{recipes}:
\begin{equation}
\rho = \frac{\sum_t (signal(t))^2}{\sum_t (noise(t))^2}.
\end{equation}
\begin{table}[h]
\caption{Magnitude $\phi$ of the noise required to reduce the
significance ApEn+FT test to ten}\vspace{12pt}
\centerline{\begin{tabular}{|c||c|c|c|} \hline
Data Set & Range & $\phi$ & $\rho$\\
\hline
chua & 6 & 0.35 & 40\\
henon & 3 & 0.45 & 2.4 \\
\hline
\end{tabular}}
\label{tab-reduce}
\end{table}

The ApEn+FT analysis of the experimental Chua circuit began to
fail with the addition of as little as $2\frac{1}{2}\%$ noise,
while the artificial H\'{e}non map could withstand up to $42\%$
noise. However, we did not find enough comparable discussions of
the signal-to-noise to make comparisons to other methods, so we
could not decide from results like these whether our method was
superior or inferior to other methods with respect to noise
sensitivity.

\subsection{H\'{e}non map subject to a linear filter}
Experimental data is sometimes filtered even before it becomes
``raw'' data\cite{eubank}. Two examples were considered for
evenly-sampled discrete data $x_n$: the moving average (MA) filter
in Equation \ref{ma} , and the linear autoregressive (AR) filter
in Equation \ref{ar}.
\begin{eqnarray}
\bar{x}_n &= &{1 \over 2\ q + 1}\sum_{m=-q}^{q} x_{m+n}
\label{ma}\\
\hat{x}_n &= &a\ \hat{x}_{n-1} + x_n \label{ar}
\end{eqnarray}
We wanted to assess the impact of these filters on data that were
clearly governed by nonlinear dynamics. Therefore we needed data
that appeared noisy or jerky because the filters would have little
effect on smooth data, (e.g., the Mackey-Glass model) even if
they were nonlinear. We chose the data from the H\'{e}non map as
described in the subsection above, because of its apparent
susceptibility to filtering, and applied both filters, for various
values of the filter parameters $q$ and $a$, respectively. The
results are listed in Table \ref{tab-filter}. The original series
(see Figure \ref{fig-data}) has an ApEn of 0.48. The sample mean
ApEn of the surrogates of the filtered data differed very little
from that of the surrogates of the unfiltered data. However, the
results in Table \ref{tab-filter} indicate that filtering
increased the ApEn of the original data by as much as $50\%$, but
the linearity hypothesis still was significantly violated ($S
\geq 100)$.
\begin{table}[h]
\caption{ApEn for MA- and AR-filtered H\'{e}non map. ApEn = 0.48
for the unfiltered H\'{e}non time series.}\vspace{12pt}
\centerline {\begin{tabular}{|c|c||c|c|} \hline
\textbf{q} (MA) & \textbf{ApEn} & \textbf{a} (AR) & \textbf{ApEn}\\
\hline
  3 & 0.69 & 0.1 & 0.46 \\
  7 & 0.76 & 0.3 & 0.43 \\
 15 & 0.65 & 0.5 & 0.51 \\
 31 & 0.59 & 0.7 & 0.69 \\
 63 & 0.58 & 0.9 & 0.62 \\
\hline
\end{tabular}}
\label{tab-filter}
\end{table}

\subsection{H\'{e}non map tested with AAFT surrogate data}
We have said throughout that we did not regard as crucial the
particular form of the method employed to generate the surrogate
time series. Nevertheless, we observed that the amplitudes of the
FT surrogate data for the H\'{e}non map were nearly twice that of
the original data (in each direction), the most we observed for
any of the data sets, so we thought it prudent to check the
consequences of such a difference in the amplitudes by applying
the amplitude-adjusted (AAFT) method\cite{theiler} to this data.
Although the AAFT surrogates look much more like the original
data, the corresponding sample mean ApEn was found to be $1.94$,
only about 5\% lower than the sample mean ApEn ($2.03$) of the FT
surrogate data; the sample standard deviations, estimated to be
$0.006$ and $0.008$ for the AAFT and the FT surrogates,
respectively, also differed little from one another.

\subsection{Logistic map with varying parameter}
The results above show the behavior of the ApEn statistic with FT
surrogate data analysis for a wide range of real data and models.
However, none of these provided insight into the behavior of the
analysis as the regularity of the data varies. Here we considered
the logistic map in order to examine the behavior of the analysis
with a parameter that changes the regularity of the data.

\begin{figure}
\centerline{\includegraphics[width = \figwidth]{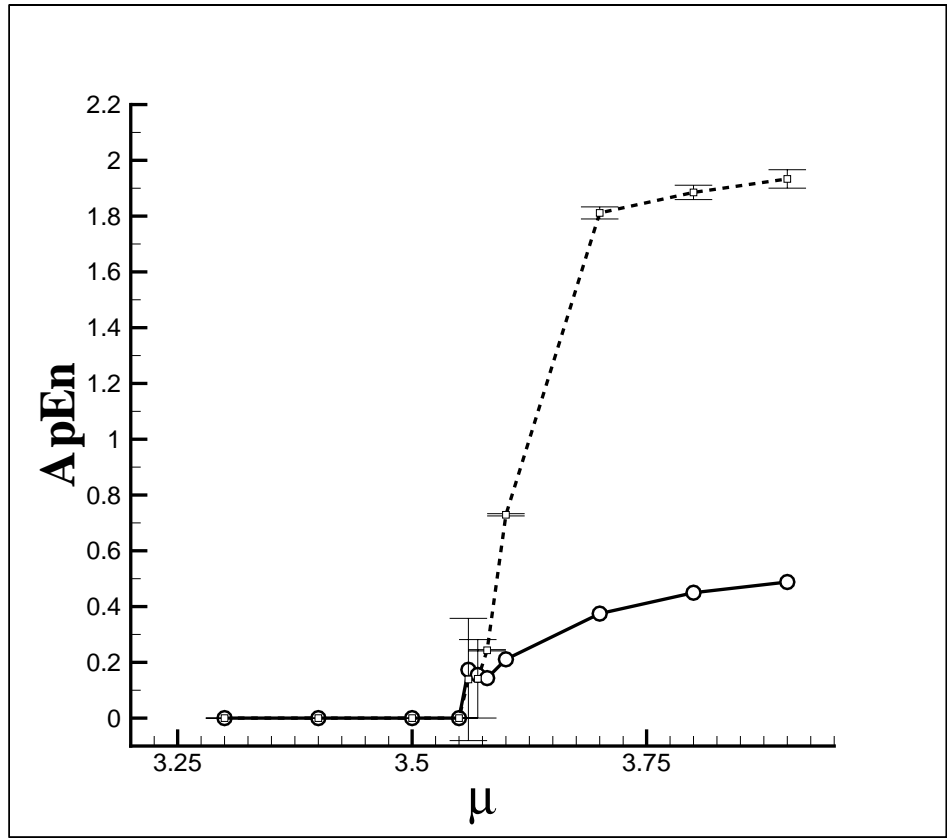}}
\caption{ApEn and FT surrogates for the logistic map for various
coupling constants. The circles and squares represent ApEn for the
original and the surrogate series, respectively. The error bars
represent the bootstrap estimate of the ($0.5\%,99.5\%$)
confidence interval for the surrogate data.} \label{fig-logistic}
\end{figure}

The logistic map is usually defined by\cite{gollub}
\begin{equation}
y_{n+1} = \mu \ y_n (1- y_n) \label{logistic}
\end{equation}
or equivalently, with a different choice of
coordinates,\cite{thermo}
\begin{eqnarray}
x_{n+1} &= &1 - \rho\ x^2_n \nonumber \\
\rho &= &{\mu\over 2}({\mu\over 2}-1)
\label{log-ulam}
\end{eqnarray}
Figure \ref{fig-logistic} shows the value of ApEn of the time
series generated from Equation \ref{logistic} and its surrogates
for various values of $\mu$. For $\mu \leq 3.55$, the dynamics are
consistent with the linearity hypothesis. ApEn begins to rise
quickly in the interval $3.55 < \mu \leq 3.70$, that contains the
Feigenbaum attractor ($\mu \approx 3.57$)\cite{thermo}, although
the comparison with the surrogates showed that the linearity
hypothesis still passed for $3.55 \leq \mu \leq 3.58$. In the
interval $3.58 < \mu < 3.60$ the linearity hypothesis finally
began to be violated, and at $\mu = 3.60$, the linearity
hypothesis was violated with $ S > 50 $. Thus this analysis
identified the critical parameter to within one percent of the
correct value.

\subsection{Kaplan-Yorke map with varying parameter}
The ApEn+FT analysis identified the transition between linear
(i.e., periodic) and nonlinear behavior with the variation of a
single parameter in the logistic map. We next examined the
transition between chaotic to random behavior with one of the
simplest of the Kaplan-Yorke family of maps\cite{thermo}, which
is a two-dimensional extension of the logistic map for $\mu = 4$
(see Equation \ref{log-ulam}):

\begin{eqnarray}
x_{n+1} &= &1 - 2x_n^2 \nonumber \\ y_{n+1} &= &k y_n + x_n
\label{KY}
\end{eqnarray}

\begin{figure}
\centerline{\includegraphics[width =
\figwidth]{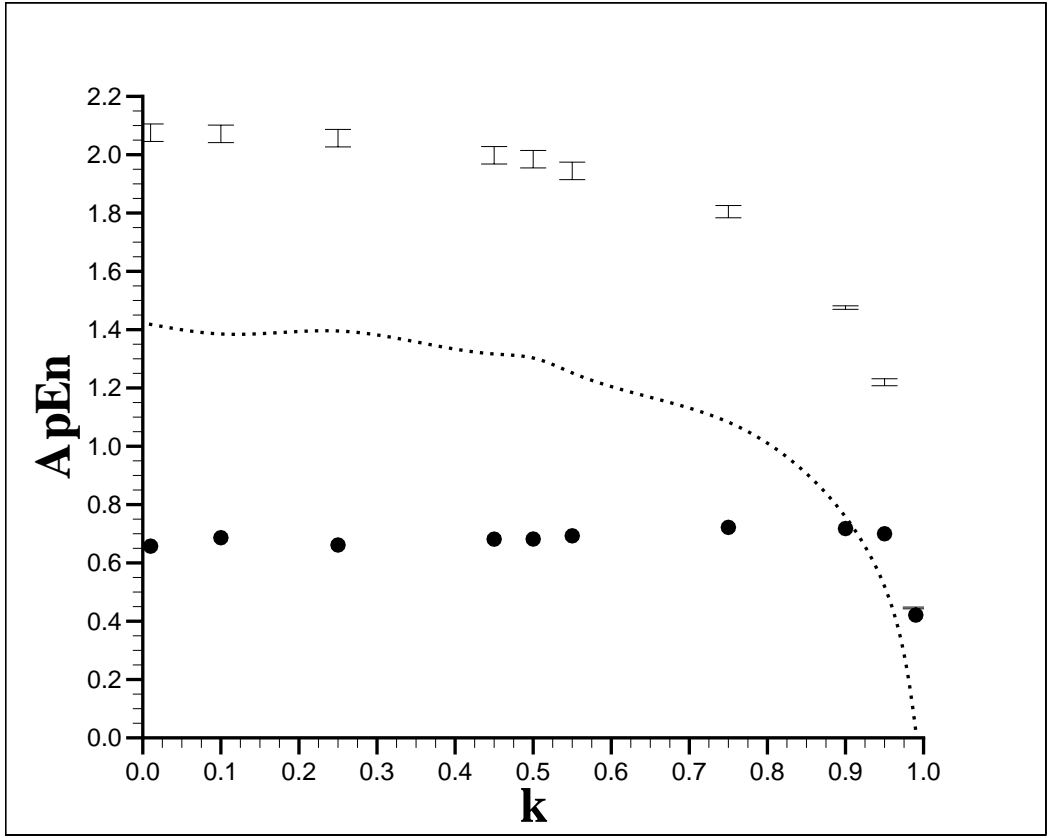}}\caption{ApEn and surrogates for the
Kaplan-Yorke map for various coupling constants. The solid circles
show the ApEn of the original time series. The height of the error
bars represents the bootstrap estimate of the ($0.1\%,99.9\%$)
confidence interval for the surrogate data. The dotted line (fit
by a spline) represents the difference between the ApEn of the
original data and the mean ApEn of the surrogate data.}
\label{fig-KY}
\end{figure}

This map has the following physical interpretation\cite{thermo}:
the $y$-coordinate describes the velocity of a unit mass particle
in a dissipative medium subject to kicks of strength $x_n$ at
times $n\tau$. For a fixed dissipation rate $\gamma$, the
dissipation parameter $k = \exp(-\gamma\tau)$. In the
small-$\tau$ limit, $\sqrt{\tau}y_n$ describes a gaussian random
process, with integral information dimension\cite{thermo}. For
moderate $\tau$ (i.e., $k \leq \frac{1}{2}$), however, Equation
\ref{KY} describes complex chaotic dynamics, with a fractional
information dimension. Figure \ref{fig-KY} shows ApEn for the
$y-$component of Equation \ref{KY}, and its surrogates, for
various $k$. Although the ApEn of the original series of $y_n$ is
flat for all but the highest values of $k$, the distance between
the ApEn of the original and the surrogate data begins to
smoothly decrease for $k > \frac{1}{2}$. The significance is at
least 100 for $k \leq 0.95$. Finally, at $k = 0.99$, the ApEn of
the original and the ApEn of the surrogate data begin to coincide
($S \approx 8$), consistent with an ARMA process. It is
interesting that the value of ApEn alone did not indicate the
transition from nonlinear to linear dynamics, since ApEn actually
decreased for the random case. Instead, the transition became
apparent only in conjunction with the surrogate data analysis.

\section{Comments and Conclusion}

The success of ApEn+FT analysis in detecting the nonlinearity in
the noiseless chaotic models (i.e., Lorenz, H\'{e}non, logistic,
Kaplan-Yorke, Mackey-Glass) was gratifying, but was neither
unique nor unexpected; indeed such success was necessary for
further investigation. The example of the Mackey-Glass model also
showed that the results could be quite sensitive to the length of
the original time series. The ability of this analysis to
distinguish between the complex chaotic behavior and the gaussian
random behavior of the Kaplan-Yorke model, although not unique,
adds to our confidence in the method. It was also encouraging to
see that even relatively severe smoothing of the H\'{e}non series
was not sufficient to obscure its inherent nonlinearity. The
ability of the ApEn+FT to detect nonlinear dynamics in the Santa
Fe Institute Data Sets A (chaotic laser) and D (turbulence) data
sets, although also not unexpected, was a more useful test because
those data sets contained some noise, and in the increments
considered here, were of intermediate length. The success of the
linearity hypothesis for Data Set E with ApEn+FT was
inconclusive, because this result might have been an accurate
assessment of that data, or it might have been a failure to
detect nonlinear dynamics in a noisy series. The strong
sensitivity of this analysis to the noise in the Chua circuit
suggests that, as with all other
techniques\cite{glass}\cite{abarbanel}\cite{kantz}, caution must
be applied when the data is both noisy and short. Direct
comparison to other methods was obstructed by the lack of
equivalent analyses in the literature, and performing such
analyses ourselves seemed beyond the scope of this paper.
Nevertheless, our goal was to empirically assess the performance
of the ApEn+FT method, and this has been obtained with the
results of the last section.

The ApEn+FT method, like other implementations of surrogate data
methods, cannot test for chaos, but only for the nonlinearity in
the dynamics of a time series that, if present, merely indicates
the potential for such a series also to be chaotic. The analysis
can say nothing further concerning possible attractors, embedding
dimension, or even if the data resulted from chaotic dynamics.
Indeed one of the most spectacular violations of the linearity
hypothesis came from the non-chaotic map of Grebogi, Ott, Pelikan
and Yorke\cite{ott}.

In addition to the problem of false positives (false success of
the linearity hypothesis), there remains the possibility of false
negatives (false violation of the linearity hypothesis) in
ApEn+FT. For example, consider the sawtooth series\cite{gollub},
with a spectrum that is poorly estimated by the straightforward
Fast Fourier Transform employed in the FT method\cite{recipes}.
Each of the resulting surrogate time series would be much more
irregular than the original time series, so that the ApEn of each
would be significantly higher than the nearly zero ApEn of the
original series, thereby yielding a false violation of the
linearity hypothesis. However, to the extent that even
complicated periodic systems might be easily recognized by visual
inspection of either the series itself or its spectrum, the
problem of false negatives might not pose the difficulties that
false positives present. If more faithful surrogate time series
were required, methods other than FT could be employed (see
Introduction), and the ApEn method could be subsequently employed
just as easily as we did here.

In conclusion, the ApEn statistic is easily implemented in the
surrogate data technique to test the hypothesis of linear
dynamics. Violations of the hypothesis were found with large
significance for a wide variety of both experimental and
computer-generated time series that were known or suspected to be
governed by nonlinear dynamics. The ApEn+FT method appears to be
useful especially for discriminating between stochastic but
linear and complex but nonlinear time series of moderate size if
the noise is not too large.

\break \textbf{Acknowledgements}

Steven Pincus provided us with valuable advice and reprints of his
papers and presentations. Karen Leighly (Oklahoma), David Peak
(Utah State), James Theiler (Los Alamos) and Jens Timmer
(Freiburg) also provided valuable scientific correspondence.

\appendix
\section{Algorithm for ApEn}
The algorithmic interpretation of ApEn, i.e., the conditional
probability that ``...runs of patterns that are close for [some
number] of observations remain close...'' was developed by Pincus
as follows\cite{pincus91}. For a sequence $\{u_{k}\}$, with
$0\leq k < N$, the $N -m$ window vectors $w(j)$ for $0\leq j < N -
m $, each of window length $m$, are defined by
\begin{equation}
w(j) \equiv [u_j,\ldots,u_{j+m}].
\end{equation}
The fraction $C_i^m(r)$ of the distances from a given window
vector $w(i)$ to all the window vectors (including itself) that
lie within a tolerance $r$ is defined by
\begin{equation}
C_i^m(r) \equiv {1 \over {N-m}}\sum_{j=0}^{N - m - 1} \Theta(r -
d[w(i),w(j)]),
\end{equation}
where $\Theta(x \geq 0) = 1$ and $\Theta(x < 0) = 0$, and the
distance $d$ is given by the $L_1$ norm, i.e.,
\begin{equation}
d[w(i),w(j)] \equiv \max_{0\leq k \leq m} |u_{i+k} - u_{j+k}|.
\end{equation}
Then ``the $C_i^m(r)$ measure, within a tolerance $r$, the
regularity of patterns similar to a given pattern of window length
$m$''.\cite{pincus95a} Collecting these ideas, Pincus defined
ApEn by
\begin{eqnarray}
\textrm{ApEn}(N,m,r) &\equiv &\Phi^m(r) - \Phi^{m+1}(r),\\
\Phi^m(r) &\equiv &\frac{1}{N-m}\sum_{i=0}^{N-m-1}\ln (C_i^m(r))
\nonumber
\end{eqnarray}

\section{ApEn and related quantities in various limits}
There were at least two ways to employ ApEn to measure regularity
in data, but only one of these was be fruitful for our
application. The differences arise from the way the parameters
$m$, $r$, and $N$ employed. Viewed in the limits $N\rightarrow
\infty$, $m\rightarrow \infty$, and $r\rightarrow 0$, Pincus
showed that the definitions above yield, at least
theoretically\cite{pincus91}, the more familiar correlation
integral $\Gamma_m(r)$ and correlation dimension $\beta$ as
follows:
\begin{eqnarray}
\Gamma_m(r) &= &\lim_{N\rightarrow
\infty}\frac{1}{N-m}\sum\limits_{i=0}^{N-m-1}
C_i^m(r),\\
\beta &= &\lim_{m\rightarrow \infty \atop r\rightarrow 0}
\frac{\ln\left(\Gamma_m(r)\right)}{\ln(r)}.
\end{eqnarray}
The theoretical bounds to ApEn in these limits are $0$ for a
perfectly regular sequence, and $\ln(B)$ for a maximally irregular
(random) sequence of base $B$ numbers. ApEn itself theoretically
converges to the Eckmann-Ruelle
entropy\cite{pincus91}\cite{pincus92}\cite{pincus95b}. Pincus has
explored the large-$N$ limit numerically for binary numbers,
taking advantage of their special structure\cite{pincus97}.
However, for most work with decimal numbers, the convergence is
so slow that the limits above are practically unobtainable,
therefore we did not attempt to employ them in this work.

\section{Connection between ApEn of binary strings and the Ising model}
\begin{figure}
\includegraphics[width = \figwidth]{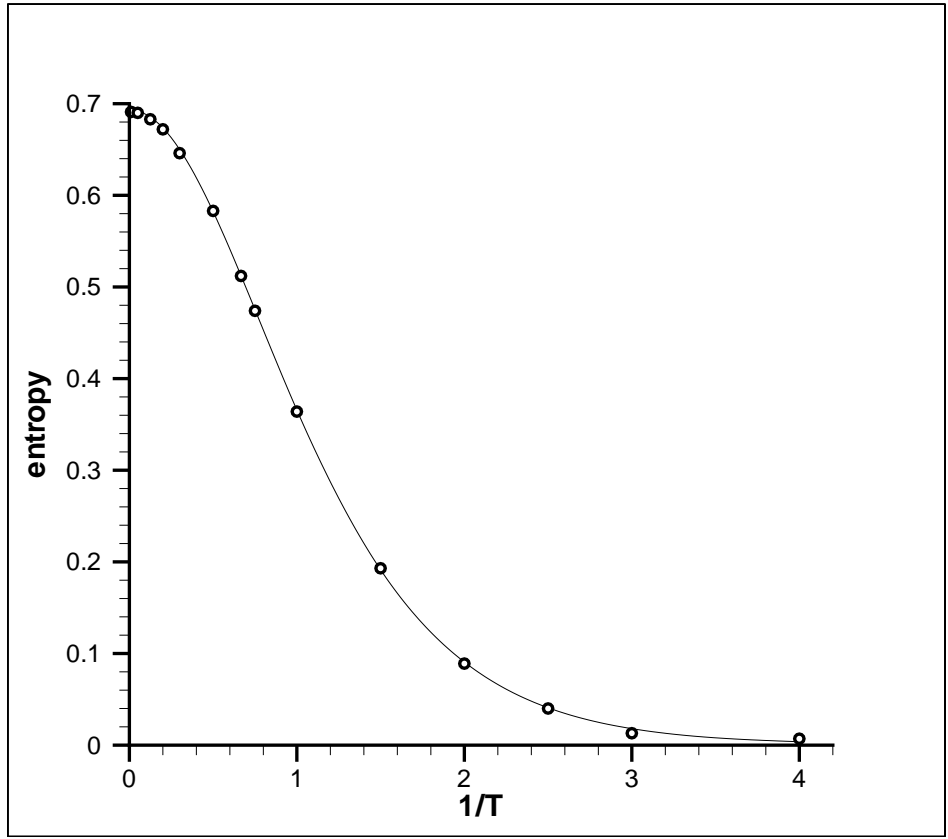} \caption{ApEn and the exact
entropy for equilibrium configurations of the Ising model for
various inverse temperatures. The solid line is the exact
thermodynamic entropy, and the circles are the calculated mean
ApEn. The estimated standard deviations are much smaller than the
circle diameter.} \label{fig-ising}
\end{figure}

As we show in Figure \ref{fig-ising}, ApEn, even with $m=2$,
essentially reproduces the thermodynamic entropy of a
one-dimensional zero-field Ising model, a two-state spin model in
which the ground state is anti-ferromagnetic, i.e., $\uparrow
\downarrow \uparrow \downarrow \uparrow \downarrow \ldots$, with
nearest-neighbor interactions on a periodic
chain\cite{thermo}\cite{thompson}\cite{chandler}. This coincidence
seems not to have been noticed before, and because it provides a
physical analogy for ApEn, we digress here briefly to include a
few details of a calculation we carried out. Although the
thermodynamics (including the entropy) are known exactly for this
model\cite{thompson}, in order to apply ApEn, we generated the
equilibrium configurations (a string of 1024 bits) from a
Monte-Carlo simulation that employed importance sampling and
periodic boundary conditions\cite{chandler}. For each temperature
we generated a total of 50 000 equilibrium configurations (binary
strings), and computed the mean ApEn($N=1024,m=2,r<1$) from 20
configurations that were separated by 2000 passes, after
discarding the first 10 000 configurations. The comparison of the
mean ApEn with the exact thermodynamic entropy is displayed in
the Figure \ref{fig-ising}. We estimated the standard deviation
of the mean ApEn to be between 0.001 and 0.002 for all
temperatures. The high temperature limit corresponds to $\ln(2)$.
There is a rough correspondence between the behavior of ApEn with
varying temperature $T$ for the Ising model and as a function of
the dissipation rate $\gamma$ for the Brownian noise model (see
Figure \ref{fig-noise}).

\break


\begin{thebibliography}{999}
\bibitem{boot} We employed 6400 bootstrap replicates of the
10-element set of ApEns. In all cases the mean and the median
practically coincided, so that we did not need nor employ the
corrections generated by more elaborate (e.g., $\rm BC_A$) methods
generally recommended by Efron and Tibshirani\cite{efron}.
\end{thebibliography}
\end{document}